\documentclass[conference]{IEEEtran}

\usepackage{amsmath,amssymb,amsfonts}
\usepackage{algorithmic}
\usepackage{graphicx}
\usepackage{textcomp}
\usepackage{dsfont}
\usepackage{tikz}

\newcommand{\E}[1]{\mathbb{E}\left[ #1 \right]}

\newcommand{\tr}[1]{\tilde{#1}}
\newcommand{\e}[1]{\bar{#1}}
\newcommand{\ex}[1]{\tilde{#1}}

\newcommand{\uB}[0]{{B}}

\begin{document}
%
\title{Predictive Network Control and Throughput Sub-Optimality of MaxWeight}

\author{
\IEEEauthorblockN{Richard Schoeffauer}
\IEEEauthorblockA{Heisenberg CIT Group\\
Free University of Berlin\\
Email: richard.schoeffauer@fu-berlin.de}
\and
\IEEEauthorblockN{Gerhard Wunder}
\IEEEauthorblockA{Heisenberg CIT Group\\
Free University of Berlin\\
Email: gerhard.wunder@fu-berlin.de}
}


%


\IEEEoverridecommandlockouts
\IEEEpubid{978–1–5386–1478–5/18/\$31.00~\copyright~2018 IEEE}

\maketitle

\begin{abstract}
We present a novel control policy, called Predictive Network Control (PNC) to control wireless communication networks on packet level, based on paradigms of Model Predictive Control (MPC). In contrast to common myopic policies, who use one step ahead prediction, PNC predicts the future behavior of the system for an extended horizon, thus facilitating performance gains. We define an advanced system model in which we use a Markov chain in combination with a Bernoulli trial to model the stochastic components of the network. Furthermore, we introduce the algorithm and present two detailed simulation examples, which show general improved performance and a gain in stability region compared to the standard MaxWeight policy.
\end{abstract}


%
\IEEEpeerreviewmaketitle

\section{Introduction and Motivation}

Modern wireless networks, such as 5G, have an increasing amount of options to route packets to multiple nodes, making information flow control essential for throughput performance, e.g. to avoid bottlenecks due to cell overload, or to exploit diversity of wireless links for low latency communication \cite{pocovi_ultra}. Seminal work on wireless network control was published in \cite{tassiulas_max_weight} in 1992 and introduced the so called MaxWeight policy (MW). This policy still stands as a benchmark and was improved upon many times, e.g. in \cite{meyn_y_max_weight} and \cite{wunder_y_maxweight}. However, MW and all its deviations are of the \textit{myopic} type \cite{meyn_complex_networks}. This means, they only make decisions based on immediate \textit{next step} system changes (while using a time discrete network description which is the most convenient model due to clocked devices in reality). Although not without its flaws (e.g. large delays of single packets under low workload \cite{subramanian_draining_time}), this approach used to be reasonable due to limited amount of computational resources and high pace requirements. Espiecally in simple network topologies (e.g. boradcasting), these policies are well studied \cite{wunder_q_analysis}.
Nowadays though, with advancements in computational power, we have the option of using more mathematically ambitious algorithms to devise policies that can improve on the network behavior. Initial attempts in this direction were made in \cite{wunder_draining}, where in each time slot a draining problem is solved to minimize delay for an OFDM broadcast channel.

In this paper, we introduce a new network control policy, that we call Predictive Network Control (PNC). Herein we use paradigms from the field of Model Predictive Control (MPC) \cite{mayne_mpc_survey}, to devise a policy that predicts the system behavior for multiple steps into the future. While we gain improvements in performance, this approach additionally produces a schedule of predicted communications which seems very intriguing for implementation into Cyber Physical Systems, where control and communication merge together.

This work is dedicated to present the new policy together with the utilized system model. We omit analytical results for later publications but provide numerical results which indicate inferiority of MW, in terms of performance and overall stability.

In the next sections, we will use upper case letters for matrices ($M$) and cost functions ($J$), bold ones for sets ($\mathbb{S}$) or probability operators ($\mathbb{P}[\cdot]$) and lower case for everything else.

\section{System Model}

 \IEEEpubidadjcol
We concentrate on the fairly common case of a discrete-time, packet-level communication network, where we have an arbitrary number of entities sending information to one another \cite{meyn_complex_networks}. Additionally, information may enter or leave the whole system. If an entity receives more information than it can send, it has to store it in a buffer. The amount of information in that buffer we call the queue length $q^{(i)}_t$ (buffer $i$, time slot $t$). 

Considering all entities at once, we get the queue-vector $q_t \in \mathbb{N}^n$, $n$ being the number of all buffers. It is possible that one entity keeps multiple buffers, e.g. if the received information is of different type. This will simply result in $n$ being larger than the number of entities. Between the entities there exist so called communication \textit{links} as column vectors. E.g. the vector
$\begin{pmatrix} -1 & 1 & 0 & \dots & 0 \end{pmatrix}^T$
would send information from the first buffer (decreasing its size by $1$) to the second buffer (increasing its size by $1$).
These $m$ vectors collected side by side form the system-matrix $B_t$. 
We can influence the network by deciding, which links are to be activated at the current time. For that, we use the binary input- or control-vector $u_t \in \{ 0,1 \}^m$.
The system evolution can then be expressed as
\begin{equation}
\label{eq::sys_evo}
q_{t+1} = q_t + B_t u_t + a_t
\end{equation}
where $B_t \in \mathbb{Z}^{n\times m}$ is discrete valued.
Information leaves the system, when columns are activated, that possess  
more negative than positive entries. This represents the information reaching its intended destination, which is the goal of any controller.
In contrast, the arrival-vector
$a_t \in \mathbb{N}^n$
represents an information influx by simply increasing $q_t$ (usually in a stochastic way).
Its time average $\e{a} = \mathbb{E} \left[a \right] \in \mathbb{R}_+^n$ we call the arrival-rate, which is pivotal for system stability.

If every possible communication link (column of $B_t$) could be activated at the same time, there would be no need to steer the system. However, in reality some of these \textit{links} might not be available at the same time, because they share the same communication \textit{channels} (which have a limited capacity). These disjunct links can be encoded in the constituency-matrix $C$ which limits the control decisions through $C u_t \leq \mathds{1}$ (with $\mathds{1}$ being the vector of ones, with appropriate dimension). This means, that the controller is usually left with the decision of which information to route first.
While this system model is mostly in alignment to the usual one \cite{tassiulas_max_weight}, we now introduce some novel features, that aim to model especially wireless communication more accurately.

(F1) we allow our links (columns of $B_t$) to have discrete values (in contrast to the usual binary ones). This allows information of different type to have different size or different importance since high entries in a communication link will mean faster transportation.

(F2) we allow each column of $B_t$ to have multiple positive and negative entries and define that we can \textit{not} activate a link, when (due to the system evolution) at least one entry of $q_t$ would become negative. The second part is equal to the control restriction
\begin{equation}
\label{eq::u_set}
u_t \in \mathbb{U}_t := \big \{
u_t \in \{0,1\}^m \ : \ Cu_t \leq \mathds{1} \ , \ q_t+B_t u_t \geq 0
\big \}
\end{equation}
where both inequalities are meant element wise.
This makes it possible to couple the processing of information, or model a demand \cite[p.273]{meyn_complex_networks} by forcing certain buffers to only be depletable together.
Especially the last inequality in \eqref{eq::u_set} deviates from the standard models used in \cite{tassiulas_max_weight} and \cite{meyn_complex_networks}, and is responsible for many results that are presented here.

(F3) as already done in \cite{tassiulas_max_weight}, we capture the short term wireless characteristics (e.g. channel fading), by introducing a diagonal weight matrix $M \in [0,1]^{m \times m}$ which encodes the success probability of an activated communication link. Hence a controller might activate some link in time slot $t$, which in the end will not influence the system due to the communication being unsuccessful. More specifically, we define that the controller has knowledge of all communication links (now encoded as columns of the matrix $\uB$) and their success probabilities (encoded in $M$). For the real system evolution however, we evaluate $M$ by performing Bernoulli trials (coin tosses) on its individual entries. Specifically, $\mathbb{B}[\xi] = 1$ w.p. $\xi$ and 0 otherwise for any $\xi \in [0,1]$. Using this operator on matrices means using it separately on its individual entries. We then obtain $B_t$ from \eqref{eq::sys_evo} through multiplication as $B_t = \uB \, \mathbb{B} \left[ M \right]$. As a result, some columns of $B_t$ are set to $0$, whereas others are as in $\uB$, independent of the decisions made by the controller.

(F4) we capture long term wireless characteristics (e.g. entities leaving the entire system), by having a whole set of probability matrices
$\left \{ M_i \right \}, i = 1\dots p$ and choosing one of these through an underlying discrete time Markov chain (DTMC) $\sigma_t = \mathbb{M}(\{1 \dots p\},P,\sigma_0)$ on its index set $\{ 1 \dots p \}$ with left side transition matrix $P$ and initial state $\sigma_0$. In each time slot this DTMC will dictate, which $M_i$ to use for the description in (F3). We can therefore write 
$B_t = \uB \, \mathbb{B} \left[ M_{\sigma_t} \right]$.
This setup is tightly related to \textit{Discrete-Time Markov Jump Linear Systems} \cite{costa_jump_markov}.

\section{Predictive Network Control (PNC)}
\label{sec::PNC}

We now introduce our new control policy, called Predictive Network Control (PNC). Like for any policy, the purpose of PNC is to let the system process as much information per time as possible, i.e. to escort it out of the system.

As mentioned, PNC is inspired by the paradigms of Model Predictive Control (MPC) \cite{mayne_mpc_survey}. Therefore we first define a cost function $J(\cdot)$, which assigns a cost to a whole trajectory of control-vectors
$\tr{u}_{t}= \begin{pmatrix}
u_{t}^T & u_{t+1}^T & \dots & u_{t+H-1}^T
\end{pmatrix}^T$. 
We call the length of the trajectory the \textit{prediction horizon} $H$. Going on, we calculate the minimizing trajectory of input-vectors $\tr{u}_t^*$ and intuitively apply its first vector $u_t^*$ to our system to advance to the next time slot $t+1$. However once there, we do \textit{not} apply the second vector of the optimal trajectory, which would be $u^*_{t+1}$, since it is outdated. We instead repeat the whole process (minimizing, applying the first input-vector of the optimal trajectory) and thus obtain an implicit feedback control law.

In what follows, we will assume that the current time slot is $t=0$.
We choose the cost $J(\cdot)$ to be of quadratic form
\begin{equation}
\label{eq::cost_function}
J(\tr{u}_0, q_0, \sigma_0) = \E{
\sum_{i=1}^{H} q_{i}^T Q q_{i} + u_{i-1}^T R u_{i-1} \ \middle| \ q_0,\sigma_0
}
\end{equation}
with $Q$ and $R$ being symmetric, positive definite matrices. 
Naturally, we do not have a choice but to work with the expectation $\E{\cdot}$ of future queue-vectors due to the stochastic system evolution. For this cost function, we can find the optimal, minimizing control $\tr{u}_0^*$ by transforming the problem into a standard quadratic program. The rest of this section shows how this can be accomplished (despite of the stochastics in the system evolution).

To handle the DTMC,  it helps to define the expanded versions of $\sigma_t$, $P$, and $\uB$ as
\begin{equation}
\ex{e}_{\sigma_t} = e_{\sigma_t} \otimes I_n
\qquad
\ex{P} = P \otimes I_n
\qquad
\ex{B} = \begin{pmatrix}
\uB M_1 \\ \vdots \\ \uB M_p
\end{pmatrix}
\end{equation}
where $e_{\sigma_t}$ is the flag-vector, corresponding to $\sigma_t$ (which is $1$ at the $(\sigma_t)$-th element and $0$ everywhere else).
With this we can express the expected future system-matrix given an initial Markov-state as
\begin{equation}
\label{eq::B_moment_1}
\e{B}_t(\sigma_0) := 
\E{B_t \ \middle| \ \sigma_0} = 
\left( \ex{P}^t \ex{e}_{\sigma_0} \right)^T \ex{B}
\end{equation}
It follows that the expected queue-vector becomes
\begin{equation}
\E{q_t \ \middle| \ q_0,\sigma_0} = q_0 + \sum_{i=0}^{t-1} \e{B}_i(\sigma_0) u_i  + t \e{a}
\end{equation}
Substituting this into the cost function yields three cost terms, which are constant, linear and quadratic with respect to our control:
\begin{equation*}
J(\tr{u}_0,q_0,\sigma_0) = 
J_{\text{c}}(q_0,\sigma_0) +
J_{\text{l}}(q_0,\sigma_0)\tr{u}_0 + 
\tr{u}_0^T J_{\text{q}}(q_0,\sigma_0)\tr{u}_0
\end{equation*}
The first term is independent of $u_t$ and therefore without concern to us.
The linear term can be expressed with \eqref{eq::B_moment_1} as
\begin{equation}
\begin{aligned}
J_{\text{l}}(q_0,\sigma_0)
=&
q_0^T 2 Q
\begin{pmatrix}
\left( H-0 \right) \e{B}_0^T(\sigma_0) \\
\left( H-1 \right) \e{B}_1^T(\sigma_0) \\
\vdots \\
\left( H-H \right) \e{B}_{H-1}^T(\sigma_0)
\end{pmatrix}^T
\\
+&
\e{a}^T Q 
\begin{pmatrix}
\left( H+1 \right) \left( H-0 \right) \e{B}_0^T(\sigma_0) \\
\left( H+2 \right) \left( H-1 \right) \e{B}_1^T(\sigma_0) \\
\vdots \\
\left( H+H \right) \left( \, 1 \, \right) \e{B}_{H-1}^T(\sigma_0)
\end{pmatrix}^T
\end{aligned}
\end{equation}
For the quadratic cost term, we need an expression for the expected square of the system-matrices. It is easy to verify that for $k \geq l$
\begin{equation}
\label{eq::expect_quad_markov}
\begin{aligned}
\bar{\bar{B}}_{k,l} (\sigma_0) :=&
\E{B_k^TQB_l \ \middle| \ \sigma_0}
\\ =&
\left(
\ex{P}^l \ex{e}_{\sigma_0}
\right)^T
\begin{pmatrix}
\ex{B}^T \ex{P}^{k-l} \ex{e}_{1} Q \uB M_1
\\
\ex{B}^T \ex{P}^{k-l} \ex{e}_{2} Q \uB M_2
\\
\vdots 
\\
\ex{B}^T \ex{P}^{k-l} \ex{e}_{p} Q \uB M_p
\end{pmatrix}
\end{aligned}
\end{equation}
It follows that
\begin{equation}
\begin{gathered}
J_{\text{q}}(q_0, \sigma_0) = J_{\text{q}}( \sigma_0) = (I_H \otimes R)+
\\
\begin{pmatrix}
H \cdot \bar{\bar{B}}_{0,0}(\sigma_0) 
& 
\left( H-1 \right) \cdot \bar{\bar{B}}_{0,1}(\sigma_0) 
& \dots
\\[1ex]
\left( H-1 \right) \cdot \bar{\bar{B}}_{1,0}(\sigma_0) 
& \left( H-1 \right) \cdot \bar{\bar{B}}_{1,1}(\sigma_0) 
& \dots
\\
\vdots & \vdots & \ddots
\end{pmatrix}
\end{gathered}
\end{equation}
Furthermore, we need to handle the following 3 constraints

(i) the binarity of $\tr{u}_0$
\begin{equation}
\label{eq::U}
\tr{u}_0 \in \{0,1\}^{m \cdot H} =: \mathbb{U}
\end{equation}

(ii) the constituency of $\tr{u}_0$
\begin{equation}
\label{eq::Cu}
\left[ I_H \otimes C \right] \tr{u}_0 \leq \mathds{1}
\end{equation}

(iii) the constraint of $q_t \in \mathbb{N}^n$. Here, the discreteness is provided by the system model \eqref{eq::sys_evo}. However, there are several ways to translate the positiveness into the optimization. Simply forcing $q_t \geq 0$ for $t=1 \dots H$, resulting in $H$ so called \textit{hard constraints}, is most conservative and neglects any information on the arrival rate (setting it to the worst case, which is 0). Indeed, it would suffice to only force positiveness for the first evolution, $q_1 \geq 0$, since this alone would already guarantee overall positiveness due to the repeated application of the optimization. The rest of the constraints could then be reformulated as \textit{soft constraints} $\E{q_t} \geq 0$ for $t = 2 \dots H$. We suggest an adjustable approach, depending on the variance of the arrival. The more evenly the arrival, the more soft constraints should be used. This should improve performance due to a better prediction of the future states of the network. Using only one hard constraint for the first two steps yields the following constraint
\begin{equation}
\underbrace{
\begin{pmatrix}
B^- & 0 & 0 & \dots \\
B & B^- & 0 & \dots  \\
\e{B}_0 & \e{B}_1 & \e{B}_2^- \\
\vdots & \vdots &  & \ddots \\
\e{B}_0 & \e{B}_1 & \e{B}_2
& \dots & \e{B}_{H-1}^-
\end{pmatrix}_{\sigma_0}
}_{\displaystyle D(\sigma_0)}
\tr{u}_0
\leq
\underbrace{
\begin{pmatrix}
q_0 \\ q_0 \\ q_0 + 3 \e{a} \\ \vdots \\ q_0 + H\e{a}
\end{pmatrix}
}_{\displaystyle d(q_0,\e{a})}
\end{equation}
Multiple hard constraints up to time slot $\tau$ can be implemented by replacing any expected system-matrices matrices with $B$ on the lhs and removing any $\e{a}$ term on the rhs of the constraint up until row $\tau$. Note that the last matrix in each row is denoted with $(\cdot)^-$. This operator transforms the matrix by setting every positive entry of it to $0$. Let $A$ be any real valued matrix and $a_{ij}$ its entries, then $a_{ij}^- := \min \{ a_{ij} , 0 \}$.
This is necessary to forbid the system to route a single packet of information through multiple queues in a single time slot.

Finally, we can define the PNC as the policy that in each time slot solves the binary quadratic program
\begin{equation}
\label{eq::PNC_program}
\begin{gathered}
\tr{u}_t^* = \arg
\min_{\tr{u}_t} \
  J_{\text{l}}(q_t,\sigma_t) \tr{u}_t
+ 
\tr{u}^T_t
J_{\text{q}}(\sigma_t) 
\tr{u}_t
\\
\text{s.t.}
\\
\tr{u}_t \in \mathbb{U}, 
\quad
\left[ I_H \otimes C \right] \tr{u}_t \leq \mathds{1}, 
\quad
D(\sigma_t) \tr{u}_t \leq d(q_t,\e{a})
\end{gathered}
\end{equation}
and initializes the first optimal control $u_t^*$ from its solution. We implicitly assume, that the Markov-state $\sigma_t$ is known together with all other used parameters.

\section{Simulations}

In what follows, we showcase the behavior of two exemplary networks, when controlled by the PNC policy. We compare it directly with the MW policy, which as mentioned is often used as a benchmark. Note however, that especially feature (F2) makes MW deviate from its usual throughput optimal behavior. Also, since the arrival has stochastic character, each simulation yields slightly different results. The here presented graphs are therefore only representatives which, to the best of our knowledge, do showcase the usual behavior of the quantity in question.

Note, that PNC, as defined in section \ref{sec::PNC}, uses a binary quadratic optimization over a control variable of potentially high dimension (depending on horizon $H$). Whereas MW uses binary linear optimization. Thus any gain in performance has to be set into relation to the additional computational cost. Having this in mind, we developed a modified version of PNC, called linear PNC (L-PNC), which does neglect the quadratic term in the optimization (setting $J_{\text{q}}(\sigma_t) = 0$ in \eqref{eq::PNC_program}). To separate both versions, we will from now on call the standard PNC (as introduced in section \ref{sec::PNC}) quadratic PNC (Q-PNC). Indeed, all simulations show little to no change in performance when using L-PNC instead of Q-PNC, though this is still subject to research. Finally, since MW does not consider any direct cost contributed to the control-vector, we choose $Q= I_n$ and $R = 0$ to be able to compare the policies.

\subsection{Generic Example}

\begin{figure}
    \centering
    \includegraphics{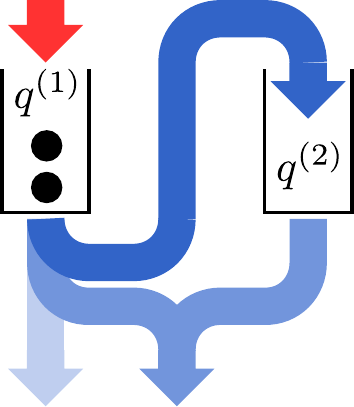}
    \caption{Sketch with Arrivals (red) and Links (blue) of generic Example}
    	\label{fig::generic_example_03}
\end{figure}

We first consider an example, specifically constructed to showcase the mechanism through which PNC dominates MW. This example does \textit{not} need any probability weights, as introduced in (F3) and (F4), but only makes use of (F1) and (F2). Specifically we look at two queues, $q_t^{(1)}$ and $q_t^{(2)}$, who are subject to arrival rates $\e{a}^{(1)}$ and $0$, respectively. As shown in Figure \ref{fig::generic_example_03}, the controller is in every time slot presented with the same three \textit{mutually exclusive} options (links), decoded as columns in the constant system-matrix
\begin{equation}
B_t = \begin{pmatrix}
-2 & -1 & -5 \\
0 & 1 & -1
\end{pmatrix}
\end{equation}
The first column directly decreases $q_t^{(1)}$ while not changing $q_t^{(2)}$. This can be interpreted as the data of $q_t^{(1)}$ being processed and afterwards leaving the system. The second column also decreases $q_t^{(1)}$ for the price of increasing $q_t^{(2)}$ which models a transmission from $q_t^{(1)}$ to $q_t^{(2)}$. The third column allows the controller to heavily decrease $q_t^{(1)}$; however according to (F2) it can only take this action if $q_t^{(2)}$ is nonempty. An interpretation could be, that the parallel processing of the information is extremely beneficial (e.g. due to a lack of storage).

For this specific case, we designed link 1 in such a way that MW will almost always prefer it compared to link 2. We say that link 1 \textit{dominates} link 2 under the MW policy. As a consequence, MW will never be able to use link 3 (since there will not be any information in $q_t^{(2)}$). Hence the maximal arrival rate that MW can handle is $\e{a}^{(1)}=-B_t^{(1,1)}=2$.

\begin{figure}
    \centering
    \includegraphics[scale=1]{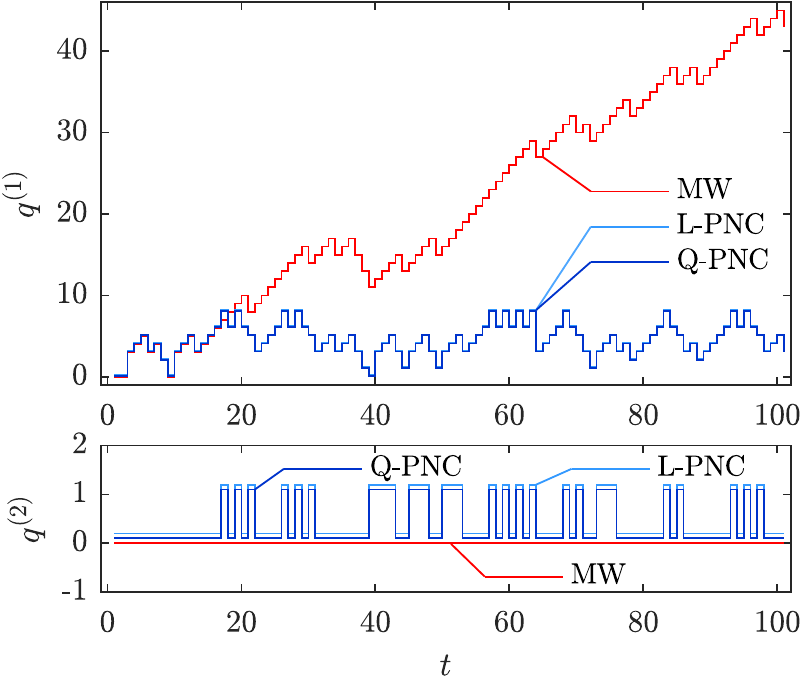}
    \caption{Instability of MW for generic Example}
    \label{fig::generic_example_01}
\end{figure}

A superior strategy would be to switch periodically between link 2 and 3, enabling the system to be stable under a maximal arrival rate of 
$\e{a}^{(1)}=-\frac{1}{2} B^{(1,2)} -\frac{1}{2} B^{(1,3)}=3 $. PNC, with a horizon of at least $H=2$ indeed follows this strategy when possible (the stochastical character of $a_t$ may prevent it from time to time). Through this behavior, PNC does stabilize the Network for a wider set of arrival rates. Figure \ref{fig::generic_example_01} compares the queue states over the first $100$ time slots for the mentioned policies. In this case, we chose $\e{a}^{(1)} = 2.4$ where we simulated $a_t$ as a Bernoulli trial (coin toss) with probability of $0.8$ and a weight of $3$. One can clearly see the growing queue length under the MW policy, whereas any of the PNC policies does result in a stable behavior. In other words: PNC can handle a larger network load than MW.

\begin{figure}
    \centering
    \includegraphics{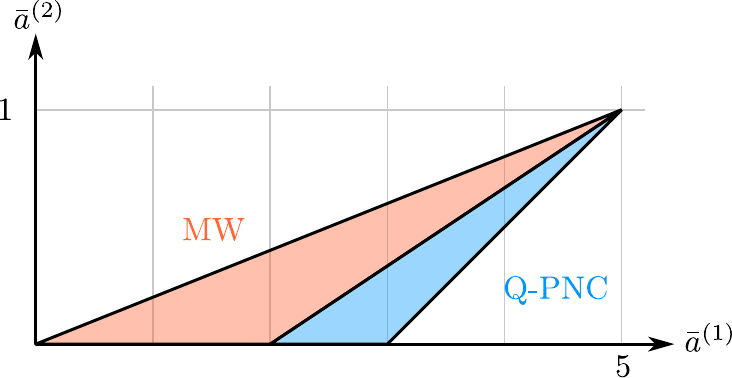}
    \caption{Stability Region for different Policies for generic Example}
    	\label{fig::generic_example_stability_region}
\end{figure}

The whole stability region for MW is shown in Figure~\ref{fig::generic_example_stability_region} as the red area. PNC, implemented with full hard constraints, does expand onto this with the blue area (but also still stabilizes the red one). For this specific example, we do not increase the stability region if we chose $H > 2$. However, there do exist such networks where the stability region increases with increasing horizon. E.g. changing the $\begin{pmatrix} -5 & -1 \end{pmatrix}$ column in the $B_t$ matrix to $\begin{pmatrix} -5 & -2 \end{pmatrix}$ would be such a network.

\subsection{Natural Example}

\begin{figure}
    \centering
    \includegraphics{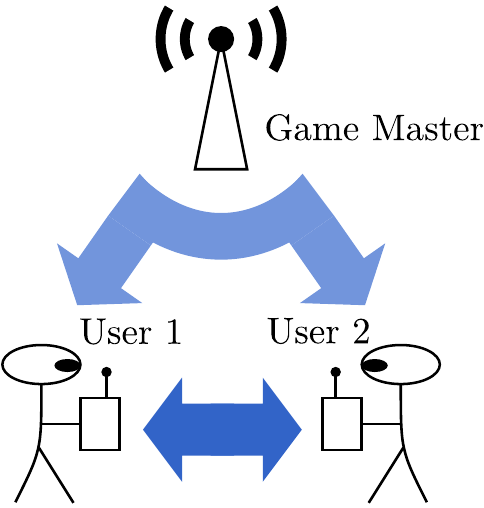}
    \caption{Real World Scenario of the natural Example}
    	\label{fig::natural_example_real_world_scenario}
\end{figure}

The second example, that is showcased in the following, is obtained by modeling a rather real world scenario. We consider two \textit{users}, interacting through a mobile game. Additionally a \textit{game master} is also needed to provide neutral information to both parties as shown in Figure \ref{fig::natural_example_real_world_scenario}. The game consists of many turns, each one progressing according to the following scheme:
The game master sends information to both users; the game waits until both users interact with this information; their inputs are communicated between each other and evaluated; the turn ends.

\begin{figure}
    \centering
    \includegraphics{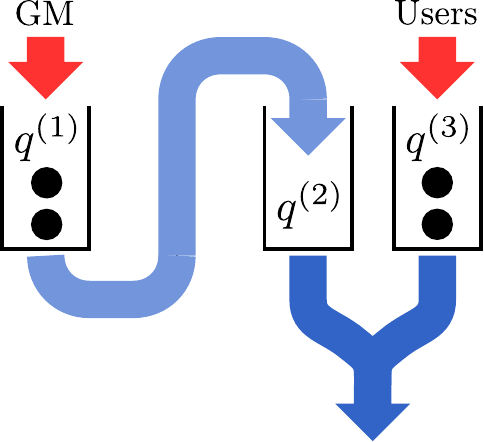}
    \caption{Sketch with Arrivals (red) and Links (blue) of the natural Example}
    	\label{fig::natural_example_model}
\end{figure}

This network can be modeled with the help of three queues: $q^{(1)}$ holds the information, the game master sends in the beginning of each turn; $q^{(2)}$ holds the information that was successfully send to both users but is not yet processed due to missing interactions by the users; $q^{(3)}$  symbolizes the inputs of the users and is increased, only when both users did interact with the information from $q^{(2)}$. Hence, information can only exit the system, when both $q^{(2)}$ and $q^{(3)}$ are non empty, i.e. only when the game master did successfully send information and \textit{both} users interacted with it.
We further define two communication channels: one between the users and one from game master to both users. We assume that both channels are mutually exclusive, so that in each time step, a policy has to decide for one of them to be inactive. Figure \ref{fig::natural_example_model} shows the model of the network with arrivals (red) and communication links (blue).
Note that $q^{(2)}$ acts as a buffer, which, when filled, allows for $q^{(3)}$ to be decreased. The superiority of PNC over MW is based on the utilization of that buffer. Through prediction of the Markov chain, PNC usually keeps this buffer at a higher level. MW on the other hand tends to use this buffer to decrease $q^{(3)}$ every time it gets the chance to do so, which is not the optimal strategy.

In this simulation example, we model the wireless characteristics according to (F3) and (F4) by introducing a \textit{good} network state, in which both communication channels are guaranteed to work, and a \textit{bad} network state, in which only the communication between the users is possible. Both are represented by the weight matrix $M_{\text{good}}$ and $M_{\text{bad}}$ respectively. We use the following set of parameters where we again model the stochastics of $a_t$ as Bernoulli trials:
\begin{equation}
\begin{gathered}
\e{a} = \begin{pmatrix}
0.5 & 0 & 0.9
\end{pmatrix}^T
,\quad
C = \begin{pmatrix}
1 & 1
\end{pmatrix}
,
\\
\uB = \begin{pmatrix}
-3 & 0 \\ 3 & -1 \\ 0 & -1
\end{pmatrix}
, \quad
P = 
\begin{pmatrix}
0.1 & 0.2 \\
0.9 & 0.8
\end{pmatrix},
\\
M_1 = M_{\text{good}} = \begin{pmatrix}
1 & 0 \\ 0 & 1
\end{pmatrix}
,\quad
M_2 = M_{\text{bad}} = \begin{pmatrix}
0 & 0 \\ 0 & 1 
\end{pmatrix}
\end{gathered}
\label{eq::parameters}
\end{equation}

\begin{figure}
    \centering
    \includegraphics[scale=1]{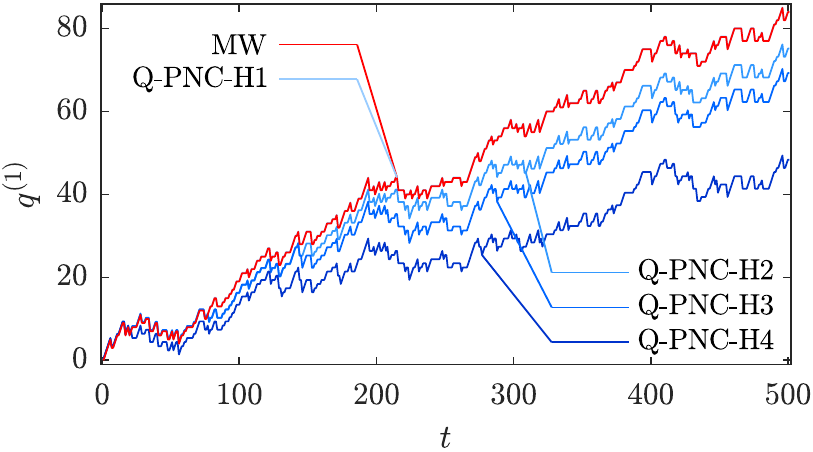}
    \caption{Performance of MW and Q-PNC for growing Horizon}
    \label{fig::natural_example_01}
\end{figure}

Figure \ref{fig::natural_example_01} shows $q^{(1)}$ over time, which is an indicator for the speed of the game. It therefore can be interpreted as a performance measure. We can see, that Q-PNC outperforms MW already for $H=2$. For $H=1$, both policies exhibit the same behavior (which as mentioned is an indication that the quadratic part of the optimization does not influence the result significantly). It should be noted, that both policies cannot stabilize the system for the specified arrival rate and that we used soft constraint for all but the first evolution of the system in the PNC algorithm.

\begin{figure}
    \centering
    \includegraphics[scale=1]{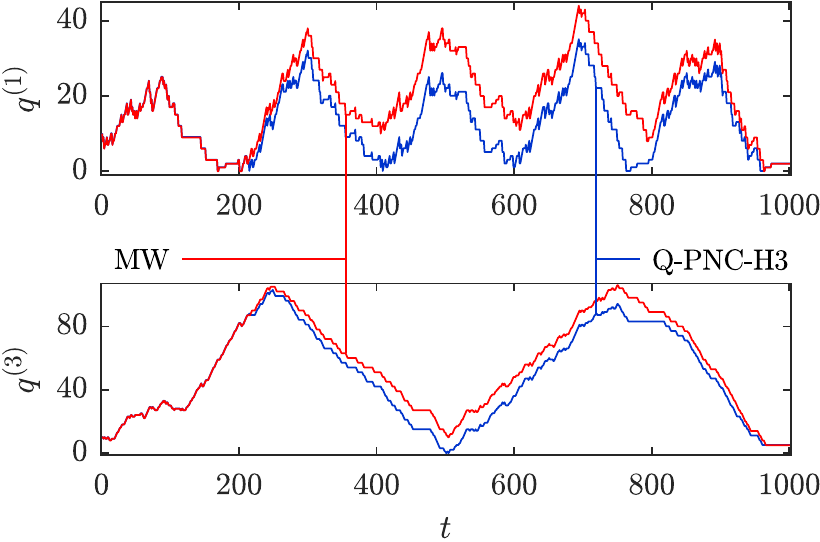}
    \caption{Performance of MW (red) and Q-PNC-H3 (blue) for stable Arrival}
    \label{fig::natural_example_alta}
\end{figure}

For a stabilizable arrival rates, Figure \ref{fig::natural_example_alta} shows the performance gains that Q-PNC with $H=3$ yields compared to MW. The average queue states are roughly about $10 \%$ smaller for Q-PNC. We omit $q^{(2)}$ which is usually higher for PNC, since it is meant to act like a storage and thus does not give any additional indication on the performance. For these graphs, we switch between periods of high and low arrivals. Specifically we used $\e{a}^{(1)} = 0.375 \pm 0.3$ and $\e{a}^{(3)} = 0.38 \pm 0.38$ for intervals $\Delta t^{(1)} = 100$ and $\Delta t^{(3)} = 250$ respectively; the arrivals were initialized as Bernoulli trials.

\section{Conclusion}

Our new control policy seems to outperform the standard MW policy when the network is based on our more general system model. Especially the gain in stability region is a huge and surprising advantage. Another conclusion, that might be overlooked when reading this paper, is the following negative indication: for rather simple networks (e.g. broadcasting scenarios), our policy does \textit{not} lead to any significant gains.
After the initial results presented here, our immediate research will evolve around the points of network classification, performance to cost trade-off and prove of stability for our new control policy.



\section*{Acknowledgment}

This paper 
is part of and thereby supported by the 
\textit{priority program SPP1914 Cyber-Physical Networking} receiving funds from the DFG
as well as the \textit{Horizon 2020 project ONE5G} (ICT-760809) receiving
funds from the European Union. The authors would like
to acknowledge the contributions of their colleagues in the
project, although the views expressed in this contribution are
those of the authors and do not necessarily represent the
project.



%

\bibliography{lib_cpn_paper}

\end{document}